\documentclass{aa}
\usepackage{graphics,ifthen}
\begin{document}
\thesaurus{ 06 (08.03.4; 08.05.3; 08.18.1; 08.19.5)}

\title{The spin-up of contracting red supergiants}

\author{
A. Heger\inst{1} 
\and 
N. Langer\inst{2}}

\institute{ 
Max-Planck-Institut f\"ur Astrophysik, Karl-Schwarzschild-Str.~1,
D--85740~Garching, Germany \\
email: ahg@mpa-garching.mpg.de
\and 
Institut f\"ur Physik, Universit\"at
Potsdam, D--14415~Potsdam, Germany\\
email: ntl@astro.physik.uni-potsdam.de
}

\offprints {A. Heger}

\date{Received  ; accepted , }

\maketitle


 
\newcommand{\Lsun}{{\mathrm{L}_{\odot}}}
\newcommand{\Msun}{{\mathrm{M}_{\odot}}}
\newcommand{\Rsun}{{\mathrm{R}_{\odot}}}
\newcommand{\cm}{{\mathrm{cm}}}
\newcommand{\km}{{\mathrm{km}}}
\newcommand{\Sec}{{\mathrm{s}}}
\newcommand{\yr}{{\mathrm{yr}}}
\newcommand{\K}{{\mathrm{K}}}
\newcommand{\erg}{{\mathrm{erg}}}
\newcommand{\Day}{{\mathrm{day}}}
\newcommand{\Mbol}{{\mathrm{M}_{\mathrm{bol}}}}
\newcommand{\Mag}{{\mathrm{mag}}}

\newcommand{\kms}{{\km\,\Sec^{-1}}}
\newcommand{\ergs}{{\erg\,\Sec}}
\newcommand{\iunit}{{\cm^2}}
\newcommand{\junit}{{\cm^2\,\Sec^{-1}}}
\newcommand{\dmsunit}{{\Msun\,\yr^{-1}}}

\newcommand{\aMLT}{{\alpha_{\mathrm{MLT}}}}
\newcommand{\aS}{{\alpha_{\mathrm{sem}}}}

\newcommand{\Prot}{{P_{\mathrm{rot}}}}
\newcommand{\tauconv}{{\tau_{\mathrm{conv}}}}
\newcommand{\Hp}{{H_{\mathrm{P}}}}
\newcommand{\vconv}{{v_{\mathrm{conv}}}}
\newcommand{\vsurf}{{v_{\mathrm{surf}}}}
\newcommand{\dt}{{\Delta\:\!\!t}}
\newcommand{\dM}{{\Delta\:\!\!M}}
\newcommand{\dJ}{{\Delta\:\!\!J}}
\newcommand{\Mdot}{{\dot{M}}}
\newcommand{\Mdotl}{{\Mdot_{\mathrm{low}}}}
\newcommand{\Menv}{{M_{\mathrm{env}}}}
\newcommand{\Jdot}{{\dot{J}}}
\newcommand{\Jenv}{{J_{\mathrm{env}}}}
\newcommand{\isurf}{{i_{\mathrm{surf}}}}
\newcommand{\ienv}{{i_{\mathrm{env}}}}
\newcommand{\Renvs}{{\cal{R}_{\mathrm{env}}}}
\newcommand{\Rs}{{R_{\star}}}
\newcommand{\jl}{{j_{\mathrm{low}}}}
\newcommand{\il}{{i_{\mathrm{low}}}}
\newcommand{\Rl}{{R_{\mathrm{low}}}}
\newcommand{\Teff}{{T_{\mathrm{eff}}}}
\newcommand{\jsurf}{{j_{\mathrm{surf}}}}
\newcommand{\oc}{{\omega_{\mathrm{c}}}}
\newcommand{\ok}{{\omega_{\mathrm{Kep}}}}
\newcommand{\N}{{\nabla}}
\newcommand{\Nrad}{{\nabla_{\mathrm{rad}}}}
\newcommand{\chil}{{\tilde{\chi}}}

\newcommand{\D}{{\mathrm{d}}}
\newcommand{\Dm}{{\D\:\!\!m}}
\newcommand{\Param}{{\;\cdot\;}}

\def\simgr{\,\hbox{\hbox{$ > $}\kern -0.8em \lower 1.0ex\hbox{$\sim$}}\,}
\def\simle{\,\hbox{\hbox{$ < $}\kern -0.8em \lower 1.0ex\hbox{$\sim$}}\,}

\newcommand{\Be}{{B[e]}}

\newcommand{\DxDy}[2]{\frac{\D #1}{\D #2}}
\newcommand{\dxdy}[2]{\frac{\partial #1}{\partial #2}}
\newcommand{\dxdycz}[3]{\left(\frac{\partial #1}{\partial #2}\right)_{#3}}
\newcommand{\av}[1]{{\langle {#1} \rangle}}
\newcommand{\avenv}[1]{{\av{#1}_{\mathrm{env}}}}

\newcommand{\jm}{{\avenv{j}}}
\newcommand{\im}{{\avenv{i}}}
\newcommand{\rse}{{\avenv{r^2}}}

\newcommand{\lSect}[1]{{\label{sec:#1}}}
\newcommand{\lFig}[1]{{\label{fig:#1}}}
\newcommand{\lEq}[1]{{\label{eq:#1}}}
\newcommand{\lTab}[1]{{\label{tab:#1}}}

\newcommand{\Tab}[1]{{Table~\ref{tab:#1}}}
\newcommand{\Fig}[1]{{Fig.~\ref{fig:#1}}}
\newcommand{\Figs}[1]{{Figs.~\ref{fig:#1}}}
\newcommand{\Figff}[1]{{\ref{fig:#1}}}
\newcommand{\Sect}[1]{{Sect.~\ref{sec:#1}}}
\newcommand{\Eq}[1]{{Eq.~(\ref{eq:#1})}}

\newboolean{ShowPicture}
\setboolean{ShowPicture}{false}
\setboolean{ShowPicture}{true}
\newcommand{\ShowPicture}[1]{\ifthenelse{\boolean{ShowPicture}}
 {{\resizebox{\hsize}{!}{\includegraphics{#1}}}}
 {{\setlength{\unitlength}{\columnwidth}\begin{picture}(1,1)
  \put(0,0){\framebox(1,1){#1}}\end{picture}}}}
\newcommand{\ShowPictureb}[1]{\ifthenelse{\boolean{ShowPicture}}
 {\center{\resizebox{0.68\hsize}{!}{\includegraphics{#1}}}}
 {\center{\setlength{\unitlength}{1.3924\columnwidth}\begin{picture}
  (1,1.707)\put(0,0){\framebox(1,1.707){#1}}\end{picture}}}}
\newcommand{\ShowPicturec}[1]{\ifthenelse{\boolean{ShowPicture}}
 {{\resizebox{\hsize}{!}{\includegraphics{#1}}}}
 {{\setlength{\unitlength}{\columnwidth}\begin{picture}(1,2.0)
  \put(0,0){\framebox(1,2.0){#1}}\end{picture}}}}

\newcommand{\mso}{\Msun}
\newcommand{\ra}{\to}

\hyphenation{pro-cess-es}


\begin{abstract}

We report on a mechanism which may lead to a spin-up of the surface of
a rotating single star leaving the Hayashi line, which is much
stronger than the spin-up expected from the mere contraction of the
star.

By analyzing rigidly rotating, convective stellar envelopes, we
qualitatively work out the mechanism through which these envelopes may
be spun up or down by mass loss through their lower or upper boundary,
respectively.  We find that the first case describes the situation in
retreating convective envelopes, which tend to retain most of the
angular momentum while becoming less massive, thereby increasing the
specific angular momentum in the convection zone and thus in the
layers close to the stellar surface.  We explore the spin-up mechanism
quantitatively in a stellar evolution calculation of a rotating
$12\,\Msun$ star, which is found to be spun up to critical rotation
after leaving the red supergiant branch.

We discuss implications of this spin-up for the circumstellar matter
around several types of stars, i.e., post-AGB stars, {\Be} stars,
pre-main sequence stars, and, in particular, the progenitor of
Supernova 1987A.


\end{abstract}


\section{Introduction}
\lSect{intro}

The circumstellar matter around many stars shows a remarkable axial
symmetry.  Famous examples comprise Supernova~1987A (Plait et
al. 1995; Burrows et al. 1995), the Homunculus nebula around
$\eta$~Carina and other nebulae around so called Luminous Blue
Variables (Nota et al. 1995), and many planetary nebulae (Schwarz et
al. 1992).  A less spectacular example are {\Be} stars, blue
supergiants showing properties which might be well explained by a
circumstellar disk (Gummersbach et al. 1995; Zickgraf et al. 1996).
Many of these axisymmetric structures have been explained in terms of
interacting winds of rotating stars (cf. Martin \& Arnett 1995; Langer
et al. 1998; Garc\'{\i}a-Segura et al. 1998), which may be
axisymmetric when the stars rotate with a considerable fraction of the
break-up rate (Ignace et al. 1996; Owocki et al. 1996).  However, up
to now only little information is available about the evolution of the
surface rotational velocity of stars with time, in particular for
their post-main sequence phases.

Single stars which evolve into red giants or supergiants may be
subject to a significant spin-down (Endal \& Sofia 1979; Pinsonneault
et al. 1991).  Their radius increases strongly, and if the specific
angular momentum were conserved in their surface layers (which may not
be the case; see below) they would not only spin down but they would
also evolve further away from critical rotation.  Moreover, they may
lose angular momentum through a stellar wind.  Therefore, it may
appear doubtful at first whether post-red giant or supergiant single
stars can retain enough angular momentum to produce aspherical winds
due to rotation.

However, by investigating the evolution of rotating massive single
stars, we found that red supergiants, when they evolve off the Hayashi
line toward the blue part of the Hertzsprung-Russell (HR) diagram may
spin up dramatically, much stronger than expected from local angular
momentum conservation.  In the next Section, we describe the spin-up
mechanism and its critical ingredients.  In Section~3 we present the
results of evolutionary calculations for a rotating $12\,\Msun$ star,
which provides a quantitative example for the spin-up.  In Section~4
we discuss the relevance of our results for various types of stars,
and we present our conclusions in Section~5.


\section{The spin-up mechanism}
\lSect{mech}

\subsection{Assumptions}
\lSect{assume}

In this paper, we
discuss the evolution of the rotation rate of post-main sequence
stars, i.e. of stars which possess an active hydrogen burning shell
source.  This shell source converts hydrogen into helium and thus
increases the mass of the helium core with time.  However, this takes
place on the very long time scale of the thermonuclear evolution of the
star, and for the following argument we can consider the shell source
as fixed in Lagrangian mass coordinate.  

The shell source provides an entropy barrier, which separates the high
entropy envelope from the low entropy core, and it marks the location
of a strong chemical discontinuity, i.e. a place of a strong mean
molecular weight gradient. Both, the entropy and the mean molecular
weight gradient, act to strongly suppress any kind of mixing through
the hydrogen burning shell source.  This concerns chemical mixing as
well as the transport of angular momentum (e.g., Zahn 1974; Langer et
al. 1983; Meynet \& Maeder 1997).  Therefore, in the following, we
shall regard the angular momentum evolution of the hydrogen-rich
envelopes of the stars under consideration as independent of the core
evolution.

This picture is somewhat simplified, as due to the inhibition of
angular momentum transport through the shell source the post-main
sequence core contraction and envelope expansion results in a large
gradient in the angular velocity at the location of the shell source,
i.e. a large shear which may limit the inhibiting effects of the
entropy and mean molecular weight gradients.  However, with the
current formulation of shear mixing in our stellar evolution code, we
find the angular momentum transport to be insignificant
(cf. \Sect{num}). In any case, the total amount of angular momentum in
the helium core is much smaller than that in the hydrogen-rich
envelope ($\simle2\,\%$ in the $12\,\Msun$ model discussed below), so
that even if all of that would be transported out into the envelope it
would not alter its angular momentum balance much.

As a star evolves into a red supergiant, its envelope structure
changes from radiative to convective, starting at the surface.  If we
assume the envelope to be in solid body rotation initially, it can not
remain in this state without any transport of angular momentum, unless
\begin{equation}
  \dxdy{\ln i}{m} = \frac{1}{2 \pi r^3 \rho}
\lEq{homolog}
\end{equation}
remains constant with time everywhere, i.e. $r^3\rho$ remains
constant.  Here, $m$ is the Lagrangian mass coordinate, $r$ the
radius, $\rho$ the density, and $i=k^2 r^2$ is the specific moment of
inertia.  For a sphere of radius $r$ the gyration constant is $k^2=2/3$,
but even for deformed equipotential surfaces of average radius $r$ one
still finds $k^2$ to be of order unity.  The assumption that $k^2$ does
not depend on $r$ was used to derive \Eq{homolog}.  Obviously, the
above homology condition does not hold for stars which change their
mode of energy transport in the envelope, since the polytropic index
$n$ changes between $n\simeq 3$ in the radiative envelope and $n\simeq
1.5$ in the convective state.

In the following we assume that convection tends to smooth out angular
velocity gradients rather than angular momentum gradients, i.e. that
convective regions tend to be rigidly rotating.  This is certainly a
good approximation at least if the rotational period ($\Prot =
2\pi/\omega$, $\omega$ is the angular velocity) is long in comparison
to the convective time scale ($\tauconv := \Hp/\vconv$), and it may
also hold for more rapid rotation if convective blobs can be assumed
to scatter elastically (cf. Kumar et al. 1995).
Latitudinal variations of the rotational velocity as deduced from
helioseismological data for the solar convective envelope (Thompson et
al. 1996) are not taken into account in our 1D stellar evolution
calculation; however, the latitudinal averaged rotation rate of the
solar convection zone deviates by no more than $5\,\%$ from solid body
rotation (cf. e.g. Antia et al. 1997).

Although, for the considerations in \Sect{spin} we assume rigid
rotation to persist in convection zones, the necessary condition to
make the spin-up mechanism described here work is only that convection
transports angular momentum on a time scale which is short compared to
the evolutionary time scale of the star, and that it leads to a
characteristic angular momentum distribution in between the cases of
constant angular velocity and constant angular momentum.  The
efficiency of the spin-up is largest for constant angular velocity and
drops to zero for the case of constant angular momentum.  The
mechanism we present here is not restricted to convection, but any
transport of angular momentum that acts towards solid body rotation is
suitable to accomplish what we describe here.

\subsection{The spin-down of convective envelopes}
\lSect{spin}

\begin{figure*}
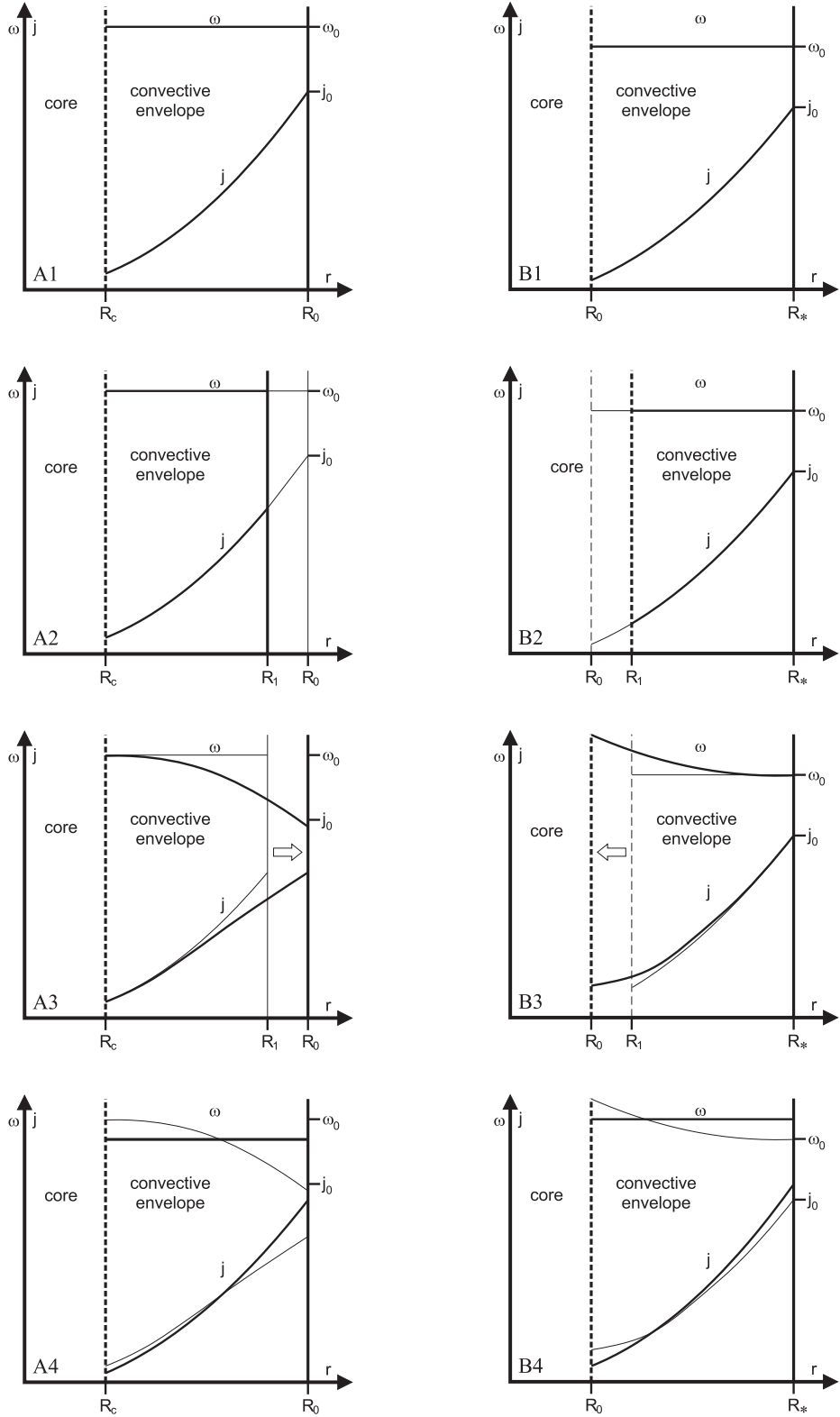

  \ShowPictureb{7414.f1}
  \caption{Mass loss from a rigidly rotating stellar envelope from the
  surface (case A: left panels) and through its lower boundary (case
  B: right panels).  For illustration, the continuous process is split
  up into three steps.  First (panels $1 \to 2$), mass is removed from
  the envelope, second, the envelope restores its original radial
  extent through expansion (panels $2 \to 3$), and third (panels $3
  \to 4$), the angular momentum is redistributed such that rigid
  rotation is restored.  It leads to spin-down (spin-up) and decrease
  (increase) of the specific angular momentum for the case of mass
  loss trough the upper (lower) boundary of the convective envelope.
  Thin lines show the state of the preceding step.}
  \lFig{panels}
\end{figure*}

The rotation frequency of a rigidly rotating convective envelope
depends on its moment of inertia and its angular momentum.  Both are
altered by mass loss from this envelope, the former by loss of
mass, the latter by the accompanied loss of angular momentum.
Additionally, the envelope's moment of inertia is also affected by
changes of its density stratification.

Here, we want to discuss two processes which can change the rotation
frequency of a rigidly rotating convective envelope without employing
global contraction or expansion of the star, but rather by mass
outflow through its upper or lower boundary.  We will show that the first
case leads to a spin-down, while the latter spins the envelope up.

The spin-down of mass losing rigidly rotating envelopes can be
understood by breaking up the continuous mass and angular momentum
loss into three discrete steps (see the left hand side of
\Fig{panels}; cf. also Langer 1998), neglecting secular changes of the
stellar structure.  Starting from a rigidly rotating envelope
extending from $R_{\mathrm{c}}$ to $R_0$ (A1), the outer part of the
convective envelope located between $R_1$ and $R_0$ is removed by
stellar mass loss within the time interval $\dt$ (A1 $\rightarrow$
A2). In the second step, for which we assume local angular momentum
conservation, the envelope re-expands to roughly its original size (A2
$\rightarrow$ A3).  This results in a slow-down of the surface
rotation and, as the layers deep inside the envelope expand less,
differential rotation below (A3).  In the third step, rigid rotation
is re-established by an upward transport of angular momentum (A3
$\rightarrow$ A4).  As this implies an averaging of the angular
velocity, it is clear from Fig~1 (A4) that now the whole convective
envelope rotates slower than at the beginning of step~1.  Obviously,
the redistribution of angular momentum towards the stellar surface
leads to an increase of the angular momentum loss rate.

The efficiency of the angular momentum loss induced by mass loss
from the surface of a rigidly rotating envelope, i.e. the amount of
angular momentum lost per unit mass lost relative to average specific
angular momentum of the envelope, is given by
\begin{equation}
  \chi:  
  =\frac{\Jdot}{\Mdot}\frac{\Menv}{\Jenv}
  =\frac{\jsurf}{\jm}
  =\frac{\isurf}{\im}
  \approx\frac{\Rs^2}{\rse}
  \;,
  \lEq{chi}
\end{equation}
where $\Menv$ and $\Jenv$ are total mass and total angular momentum of
the envelope, $\Mdot$ and $\Jdot$ are stellar mass and angular
momentum loss rate, $\jsurf$ and $\isurf$ are the specific angular
momentum and moment of inertia at the surface, respectively, and $\Rs$
is the stellar radius.  The mean value of a quantity $x$ over the
envelope is defined by
\[
  \avenv{x}:
  =\frac{1}{\Menv}\int_{M-\Menv}^{M}
  \!\!\!\!\!\!\!\!\!\!\!\!\!\!\!\!x(m)\,\Dm
  \;,
\]
where $M$ is the mass of the star.  The larger $\chi$ the more
efficient is the loss of angular momentum per unit mass lost.  A value
of $\chi>1$ corresponds to a decrease of the mean specific angular
momentum of the envelope, $\chi<1$ to an increase.  For the case of
mass loss from the surface of a rigidly rotating stellar envelope
$\chi$ is always greater than $1$.

The density stratification in the envelope determines how much angular
momentum is stored in the layers close to the surface relative to the
total angular momentum of the envelope.  An envelope structure which
holds most of the mass close to its bottom favors high angular
momentum loss rates, since this decreases $\rse$.  In red supergiants,
on the other hand, $\rse$ is rather high, since those stars store 
a large fraction of their 
mass in layers far from the stellar center, thus reducing $\chi$.

\begin{figure}
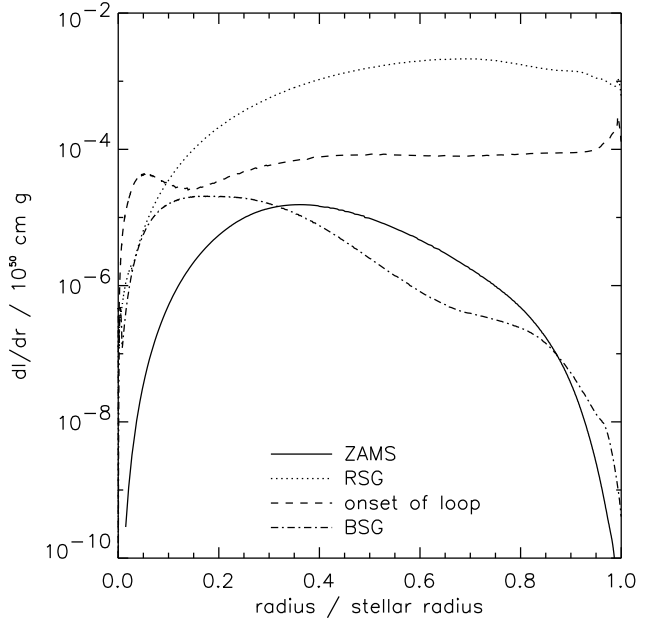

  \ShowPicture{7414.f2}
  \caption{Moment of inertia per radius $\D I(r)/\D r$ as a function
           of radius (in units of the stellar radius) for four
           different stellar models.  The solid line shows a ZAMS
           model, the dotted line a red supergiant model with an
           almost fully convective hydrogen-rich envelope, the dashed
           line a red supergiant model just before the blue loop, and
           the dash-dotted line a blue supergiant during the blue
           loop.  All models are taken from the $12\,\Msun$ sequence
           described in detail below.}
  \lFig{r-I}
\end{figure}

This is demonstrated in \Fig{r-I}, where we plotted the moment of
inertia per unit radius, $\D I(r)/\D r= 8\pi r^4 \rho/3$ for spherical
symmetry, as a function of the fractional radius for different types
of stars.  For a red supergiant model with extended convective
envelope we find the major contribution to the total moment of inertia
$I$ of the star at radii around $\sim 2/3\,\Rs$ ($\chi \approx 9/4$),
whereas for the zero-age main sequence (ZAMS) model matter at $\sim
1/3\,\Rs$ dominates the moment of inertia of the star ($\chi \approx
9$).  As main sequence stars can be approximated by rigid rotators
(cf. Zahn 1994), the whole star takes the r\^{o}le of the stellar
envelope as far as our definition of $\chi$ is concerned.  Therefore,
the considered main sequence star loses its angular momentum four
times more efficient than the red supergiant model plotted in \Fig{r-I}.

Two other cases are shown in \Fig{r-I}, i.e., a red supergiant model
shortly before its transition into the blue supergiant stage, where
$I$ is dominated by the density inversion at the upper edge of the
convective region, making the angular momentum loss from the envelope
of this star quite inefficient ($\chi \approx 1.5$), and a blue
supergiant model during central helium burning in which the moment of
inertia is concentrated even more towards the center of the star
($\chi\approx 30$) than for the ZAMS model.  Thus, if the envelopes of
blue supergiants stay close to solid body rotation --- which is
indicated by our time-dependent calculations --- they experience a
more efficient spin-down than main sequence stars.

However, since for blue supergiants the total moment of inertia $I$ is
smaller than for their progenitor main sequence stars, they may get
closer to critical rotation (cf. \Sect{phys}) if they keep their
angular momentum.  This may be in particular the case for metal poor
massive stars. Those have much smaller mass loss and therefore angular
momentum loss rates, and they can evolve from the main sequence
directly into a long-lasting blue supergiant stage without an
intermediate red supergiant phase (cf. e.g. Schaller et al.  1992); we
found such stars to obtain critical rotation as blue supergiants in
preliminary evolutionary calculations.

\subsection{The spin-up of convective envelopes}
\lSect{spun}

The spin-up of convective envelopes which decrease in mass due to an
outflow through their {\em lower} boundary is understood in a similar
way to the mass loss process discussed above. Again we want to split
up this process into tree discrete steps (see sequence B on the right
hand side of \Fig{panels}).  Starting from a rigidly rotating envelope
extending from $R_{\mathrm{0}}$ to $\Rs$ (B1), the inner part of the
convective envelope, located between $R_0$ and $R_1$, becomes
radiative within some time $\dt$ (B2).
Then the envelope re-expands to its original size. 
As local angular momentum conservation is assumed in this step, it
results in a spin-up of most of the convective envelope, which is
weaker for the layers farther out.  The outcome is a differentially
rotating envelope which spins fastest at its bottom (B3).  Since the
envelope has to go back to rigid rotation, angular momentum is
transported upward in the next step until this is achieved (B4).  We
end up with higher specific angular momentum in the whole convective
envelope compared to the initial configuration.

As can be seen in \Fig{t-r-m} below, during the evolution of a red
supergiant towards the blue part of the HR~diagram the radial extent
of the convective envelope remains about fixed while mass shells
drop out of the convective region.  If one imagines the convective
envelope to consist of moving mass elements or blobs, the spin-up
process can thus also be understood as follows.  A blob, starting
somewhere in the convective region will, as it approaches the lower
edge of the convective envelope, decrease its specific moment of
inertia and therefore has to lose angular momentum in order to remain
in solid body rotation with the whole convective region.  Angular
momentum has to be transferred to rising blobs such that also they
remain in solid body rotation.  Mass elements leaving the convective
envelope thus only remove small amounts of angular momentum from the
convective region.  Therefore, the average specific angular momentum
of the remaining convective envelope will increase and thus it spins
up.

Replacing $\jsurf$, $\isurf$ and $\Rs$ in \Eq{chi} by $\jl$, $\il$ and
$\Rl$, the specific angular momentum, the specific moment of inertia
and the radius at the lower edge of the convective envelope,
respectively, one can define an efficiency of angular momentum loss
through the lower boundary, $\chil$.  We find $\chil \ll 1$, especially
for those cases where $\chi$ is small, as e.g. for the red supergiant
envelopes under consideration here.

The total change of specific angular momentum of the envelope $\jm$ by
mass loss through the upper and lower boundary can the be written as
\begin{equation}
  \DxDy{\ln \jm}{t}
  =\frac{\Mdot\left(1-\chi\right)+\Mdotl\left(1-\chil\right)}
        {\Menv}
\;,
\lEq{dlnj}
\end{equation}
where $\Mdotl$ is the rate at which mass leaves the envelope through
its lower boundary.  For our sample calculation it is $\chil \ll 1$,
$\chi$ of order unity, and $\Mdot \ll \Mdotl$ during the red
supergiant stage preceding the evolution off the Hayashi line, so that
$\jm$ almost increases as $\Menv$ decreases,
\[
  \D\,\ln\jm/\D\,\ln\Menv\simgr-1
\;,
\]
reflecting the fact that angular momentum does not get lost
efficiently from the convective envelope through its upper nor its
lower boundary.  The approximation $\chil \ll 1$ is hardly affected by
the variation of the lower bondary radius seen in \Fig{t-r-m}.  Thus,
the convective envelope loses most of its mass but keeps a major part
of the angular momentum.

When the convective envelope gets depleted in mass and the stellar
radius decreases considerably, the global contraction of the stellar
envelope results in an additional contribution to its spin-up.  Still,
mass elements drop out of the convective region (cf. \Fig{t-r-m}), but
now the contraction leads to an increase of the rotation velocity, and
the star can reach rotation velocities of the order of the break-up
velocity (see below).  However, the contraction does not contribute to
the increase of the specific angular momentum at the surface.

Finally, we want to note that if the whole star would remain rigidly
rotating, e.g., due to the action of magnetic fields inside the star
or by more efficient shear instabilities, its spin-up would occur very
similar to the case described here.  In that case, also mass shells
from the core would transfer part of their angular momentum to layers
above, which would make the spin-up somewhat more efficient.  However,
since in a red supergiant the mass elements lose the major part of
their angular momentum to the convective envelope before they leave
it, and the amount of angular momentum contained in the core is small
anyway (cf. \Sect{assume}), the additional spin-up will be small
(cf. \Fig{t-v} before blue loop).


\section{Numerical simulations}
\lSect{num}

\subsection{Input physics}
\lSect{phys}

The stellar evolution calculations presented here are obtained with an
implicit hydrodynamic stellar evolution code (cf. Langer et al. 1988).
Convection according to the Ledoux criterion and semiconvection are
treated according to Langer et al. (1983), using a mixing length
parameter $\aMLT=1.5$ and semiconvective mixing parameter of
$\aS=0.04$ (Langer 1991a). Opacities are taken from Alexander (1994)
for the low temperature regime, and form Iglesias et al. (1996) for
higher temperatures.  The effects of rotation on the stellar structure
as well as the rotationally induced mixing processes are included as
in Pinsonneault et al. (1989), with uncertain mixing efficiencies
adjusted so as to obtain a slight chemical enrichment at the surface
of massive main sequence stars (cf. Fliegner et al. 1996).

The mass loss rate was parameterized according to Nieuwenhuijzen \& de
Jager (1990), but modified for a rotationally induced enhancement of
mass loss as the star approaches the $\Omega$-Limit (cf. Langer 1997),
i.e.
\[
  \frac{\Mdot(\omega)}{\Mdot(\omega=0)} = 
        \left(\frac{1}{1-\Omega}\right)^{0.43}
\]
(cf. Friend \& Abbott 1986) where
\[
  \Omega := \frac{\omega}{\oc}
\;,\;\;\;\;
  \oc := \sqrt{\frac{Gm}{r^3}\left(1-\Gamma\right)}
\;,\;\;\;\; 
  \Gamma := \frac{\kappa L}{4 \pi c G M}
\;.
\] 
Here, $\kappa$ is the Rosseland opacity. $\Gamma$ is not only
evaluated at the stellar surface, but its maximum value in the
radiative part of the optical depth range $\tau \in [2/3 , 100]$
(cf. Lamers 1993; Langer 1997) is used.

The quantitative result for the $\Omega$-dependence of the mass loss
rate obtained by Friend \& Abbott (1986) was questioned by Owocki et
al. (1996), who performed hydrodynamic simulations of winds of
rotating hot stars including the effect of non-radial radiation forces
and gravity-darkening in the way described by von Zeipel (1924).
However, the only crucial ingredient for the model calculations, which
is confirmed by Owocki et al. (1997), is the fact that the mass loss
rate increases strongly as the star approaches the $\Omega$-limit, so
that the star cannot exceed critical rotation, but rather loses more
mass and angular momentum (cf. also Langer 1998).

To quantify the angular momentum loss due to stellar winds, $\Jdot$,
we assume that, at any given time, the mean specific angular momentum
of the wind material leaving the star equals the specific angular
momentum averaged over the rigidly rotating, spherical stellar
surface, which we designate as $\jsurf$.

The transport of angular momentum inside the star is modeled as a
diffusive process according to
\[
  \dxdy{\omega}{t}=\frac{1}{4\pi r^2
  \rho i}\dxdy{}{r}\left[4 \pi r^2 \rho i D \dxdy{\omega}{r}\right]
  -\dot{r}\left(\dxdy{\omega}{r}+\omega\dxdy{\ln i}{r}\right)
 ,
\] 
where $D$ is the diffusion coefficient for angular momentum transport
due to convection and rotationally induced instabilities (cf. Endal \&
Sofia 1979; Pinsonneault et al. 1991), 
and the last term on the right hand side accounts for advection.  
The diffusion equation is solved
for the whole stellar interior.  In stable layers, the diffusion
coefficient is zero.  We specify boundary conditions at the stellar
surface and the stellar center which guarantee angular momentum
conservation to numerical precision.

Our prescription of angular momentum transport yields rigid rotation
when and wherever the time scale of angular momentum transport is
short compared to the local evolution time scale, no matter whether
rotationally induced mixing or convective mixing processes are
responsible for the transport.  Chaboyer \& Zahn (1992) found that
meridional circulations may lead to advection terms in the angular
momentum transport equation, which may have some influence on the
evolution of the angular momentum distribution in the stellar envelope
during the main sequence phase (cf. Talon et al. 1997).  However, the
omission of these terms has no consequences for the spin-up process
describe here, since it is dominated by the much faster angular
momentum transport due to convection.  Since the time scale of
convection is generally much smaller than the evolution time,
convective regions are mostly rigidly rotating in our models.

The effect of instabilities other than convection on the transport of
matter and angular momentum are of no relevance for the conclusions
obtained in the present paper.  E.g., mean molecular weight gradients
have no effect on the spin-up process described below since it occurs
in a retreating convection zone (see below).

Calculations performed with a version of the stellar evolution code
KEPLER (Weaver et al. 1978), which was modified to include angular
momentum, confirm the spin-up effect obtained with our code which is
described in \Sect{res}.

\subsection{Results}
\lSect{res}

We have computed stellar model sequences for different initial masses
and compositions (cf. Heger et al. 1997), but here we focus on the
calculation of a $12\,\Msun$ star of solar metallicity.  This
simulation is started on the pre-main sequence with a fully
convective, rigidly rotating, chemically homogeneous model consisting
of $28\,\%$ helium and $70\,\%$ hydrogen by mass and a distribution of
metals with relative ratios according to Grevesse \& Noels (1993).
Its initial angular momentum of $110 \,10^{50} \,\ergs$ leads to an
equatorial rotation velocity of $\sim 200\,\kms$ on the main sequence,
which is typical for these stars (cf. Fukuda 1982; Halbedel 1996).

During the main sequence evolution the star loses $25\,\%$ of its
initial angular momentum and $0.23\,\Msun$ of its envelope due to
stellar winds.  
After the end of central hydrogen burning, the star settles on the red
supergiant branch after several $10^4\,\yr$ and develops a convective
envelope of $8.3\,\Msun$.  During this phase it experiences noticeable
mass loss ($\Mdot\approx 8\,10^{-7}\,\dmsunit$), but the angular
momentum loss per unit mass lost $\Jdot/\Mdot \approx
1.7\,10^{19}\,\junit$ is lower than on the main sequence
(cf. \Sect{spin}).

\begin{figure}
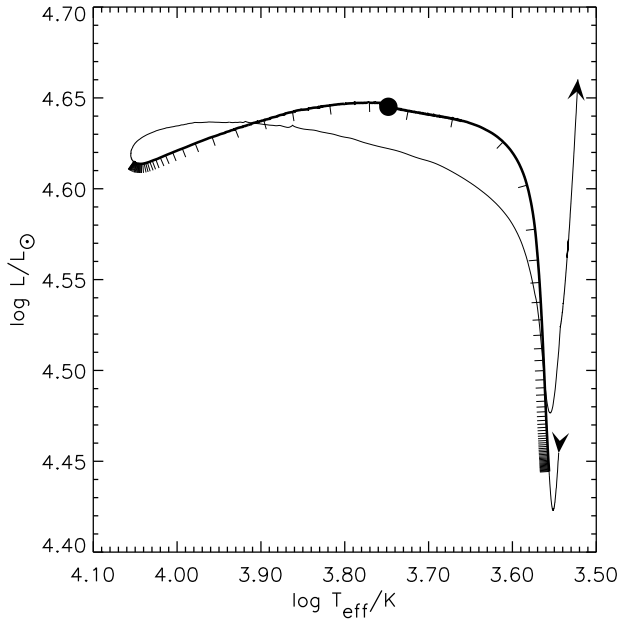

  \ShowPicture{7414.f3}
  \caption{Track of the $12\,\Msun$ star in the HR~diagram during its
  blue loop.  The two arrows indicate the direction of the evolution.
  The thick drawn part of the track corresponds to the time span shown
  in \Figs{t-xx} and \Figff{t-v}.  The star spends $1\,000\,\yr$
  between two neighboring tick marks.  The thick dot indicates where
  the star obtains its maximum mass and angular momentum loss rates;
  see the dotted line in \Fig{t-xx}.}
  \lFig{hrd}
\end{figure}

Shortly after core helium ignition, the bottom of the convective
envelope starts to slowly move up in mass.  About $4\, 10^5\,$yr later
(at a central helium mass fraction of $65\,\%$), the convective
envelope mass decreases more rapidly.  After another $\sim
5\,10^{5}\,\yr$ it reaches a value of $\sim3\,\Msun$.  Up to this
time, the star has lost $0.75\,\Msun$ as a red supergiant, and
$43\,\%$ of the angular momentum left at the end of the main sequence.

\begin{figure}
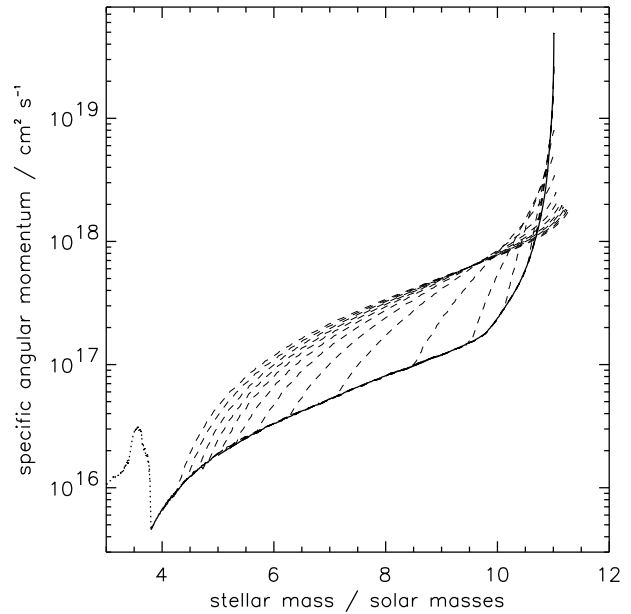

  \ShowPicture{7414.f4}
  \caption{Evolution of the internal specific angular momentum profile
  in the hydrogen-rich envelope of the $12\,\Msun$ sequence during the
  first part of core helium burning which is spent as a red
  supergiant.  Dashed lines show specific angular momentum versus mass
  coordinate in the convective part of the envelope for different
  times, from $420\,000\,\yr$ to $2\,000\,\yr$ before the red $\to$
  blue transition, with those reaching down to lower mass coordinates
  corresponding to earlier times and extending to higher mass
  coordinates, because the stellar mass decreases with time.  The
  solid line shows the angular momentum profile for those mass zones
  which have dropped out of the convective envelope, which remains
  frozen in later on.  The dotted line marks the angular momentum
  profile in the inner stellar region which is never part of the
  convective envelope.}
  \lFig{m-j}
\end{figure}

At its largest extent in mass, the convective envelope contains a
total angular momentum of $77\,10^{50}\,\ergs$, of which $17 \,10^{50}
\,\ergs$ are contained in the lower $4.5\,\Msun$.  After those layers
dropped out of the convective envelope, they have kept only $2.8
\,10^{50} \,\ergs$ whereas $44 \,10^{50}$ $\,\ergs$ remain in the
convective envelope; $30 \,10^{50} \,\ergs$ have been lost due to mass
loss.  Thus, on average the specific angular momentum of the lower
part of the convective envelope is decreased by up to a factor of
$\sim 5$, while that of the remaining convective envelope increased on
average by a factor of $1.5$, despite the angular momentum loss from
the surface (cf. \Fig{m-j}, \Eq{dlnj}).  For comparison, the helium
core contains never more than $\sim 1.5 \,10^{50} \,\ergs$.

\begin{figure}
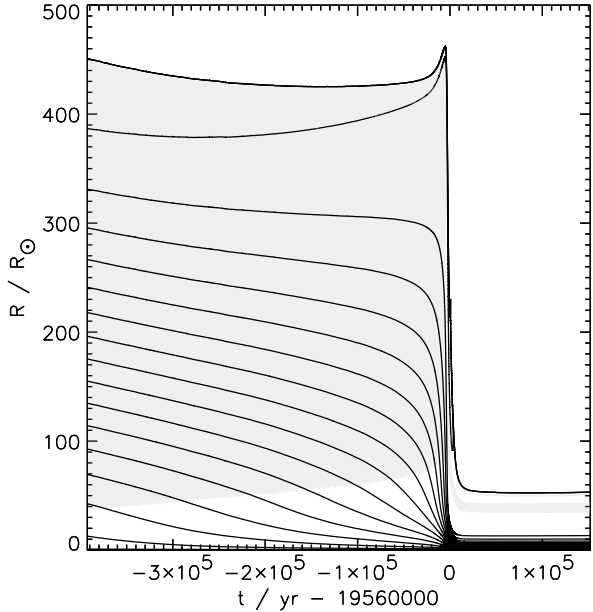

  \ShowPicture{7414.f5}
  \caption{Evolution of the radii of different mass shells as a
  function of time for a period including the transition from the red
  to the blue supergiant stage of the $12\,\Msun$ model.  The
  zero-point on the x-axis corresponds to the thick dot in \Fig{hrd}
  and corresponds roughly to the time of the red $\to$ blue
  transition.  Except for the uppermost solid line, which corresponds
  to the surface of the star, the lines trace Lagrangian mass
  coordinates.  The mass difference between the lines is $0.5\,\Msun$.
  Shading indicates convective regions.  }
  \lFig{t-r-m}
\end{figure}

\begin{figure}
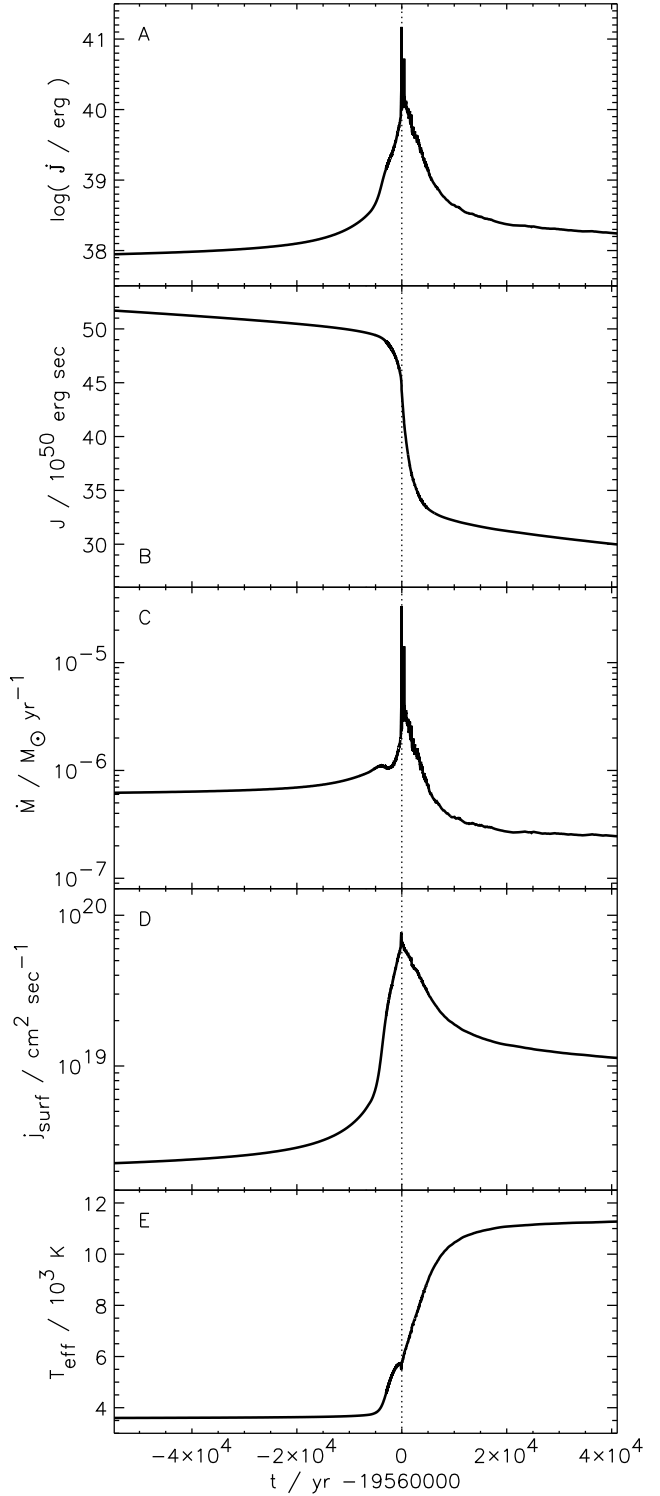

  \ShowPicturec{7414.f6}
  \caption{Evolution of characteristic stellar properties as function
  of time, during the first part of the blue loop of our $12\,\Msun$
  model (cf. \Fig{hrd}).  The time zero-point is defined as in
  \Fig{t-r-m} and is marked by the dotted line.  Displayed are: the
  angular momentum loss rate $\Jdot$ (\textbf{A}), the total angular
  momentum $J$ (\textbf{B}), the mass loss rate $\Mdot$ (\textbf{C}),
  the specific angular momentum loss rate $\jsurf=\Jdot/\Mdot$
  (\textbf{D}), and the effective temperature $\Teff$ (\textbf{E}).}
  \lFig{t-xx}
\end{figure}

At this point of evolution, the transition to the blue sets in in our
model.  Within the next $\sim 25\,000\,\yr$ another $2\,\Msun$ drop
out of the convective envelope which then comprises only about
$1\,\Msun$ but has still the full radial extent of a red supergiant.
The ensuing evolution from the Hayashi line to the blue supergiant
stage takes about $10\,000\,\yr$ (cf. \Fig{hrd}).  During this time,
the angular momentum transport to the outermost layers of the star
continues, tapping the angular momentum of those layers which drop out
of the convective envelope.

The contraction of the star by a factor of $f \approx 10$
(cf. \Fig{t-r-m}) reduces the specific moment of inertia at the
surface by $f^2 \approx 100$ and $\omega$ would be increased by the
same factor if $j$ were conserved locally.  This would imply an
increase of the equatorial rotational velocity by a factor of $\sim
10$ (cf. \Fig{t-v}, ``decoupled'').  The true increase in the
rotational velocity is one order of magnitude larger, due to the
spin-up effect described in \Sect{mech}, and it would be even larger
if the star would not arrive at the $\Omega$-limit (see \Fig{t-v}) and
lose mass and angular momentum at an enhanced rate.

For local angular momentum conservation $\Omega$ scales as $\Omega
\propto 1/\sqrt{R (1-\Gamma )}$.  A value of $\Gamma \approx
2\,10^{-3}$ is found on the red supergiant branch in our calculation,
and $\Gamma \approx 0.4$ for the blue supergiant stage.  This by
itself would increase $\Omega$ by a factor of $\sim 3.9$ during the
red $\to$ blue transition.  Actually, $\Omega$ changes from $\approx
0.01$ at the red supergiant branch at the beginning of central helium
burning to $\Omega\approx 1$ --- to critical rotation --- at the red
$\to$ blue transition, i.e. by a factor $\sim 100$, despite
significant mass and angular momentum loss.

At an effective temperature of $\sim6\,000\,\K$ the
Edd\-ing\-ton\--fac\-tor $\Gamma$ at the surface (as defined above)
rises from a few times $10^{-3}$ to $\sim0.4$, mainly due to an
increase in the opacity; the luminosity remains about constant.
Around $40\,\%$ of the angular momentum of the star are then
concentrated in the upper $0.01\,\Msun$.  Since $\Omega$ becomes close
to~1, the mass loss rate rises to values as high as several
$10^{-5}\,\dmsunit$ (cf. \Fig{t-xx}) in order to keep the star below
critical rotation, with the result that these layer are lost within a
few $1\,000\,\yr$.  This is by far the most dramatic loss of angular
momentum, i.e. the highest value of $\Jdot$, the star ever experiences.
The specific angular momentum loss $\Jdot/\Mdot$
reaches a peak value of $7.7\,10^{19}\,\junit$ (cf. \Fig{t-xx}).

\begin{figure}
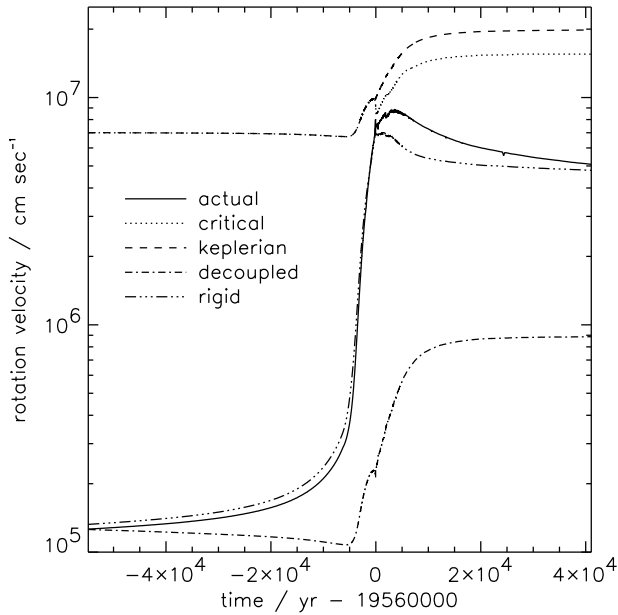

  \ShowPicture{7414.f7}
  \caption{Equatorial rotation velocity as function of time (solid
  line) compared to the Keplerian (dashed line) and the critical
  (dotted line) rotation rate; the latter two are different by the
  factor $\sqrt{1-\Gamma}$.  During the red supergiant phase it is
  $\Gamma \ll 1$ and the two lines coincide, while during the blue
  supergiant phase $\Gamma$ rises to $0.4$.  The dash-dotted line
  shows the evolution of the surface rotation rate if there were no
  angular momentum transport in the convective envelope, and the
  dash-triple-dotted line shows how the surface rotation rate would
  evolve if the whole star would maintain rigid rotation.}
  \lFig{t-v}
\end{figure}

After arriving at the blue supergiant stage, the star still
experiences a high angular momentum loss rate for some time, since the
major part of the star's angular momentum is still concentrated in the
vicinity of the surface.  Interestingly, the star now even spins
faster than when rigid rotation of the whole star were assumed
(cf. \Fig{t-v}), because within the blue supergiant's radiative
envelope the angular momentum is not transported downward efficiently.
However, due to mass loss, this deviation does not persist long.

Due to the long duration of the blue supergiant phase, several
$10^5\,\yr$ in comparison to the $10^4\,\yr$ the red $\to$ blue
transition takes, the total angular momentum $J$ is reduced by another
factor $\sim 3$ although both, the mass and the angular momentum loss
rate, become smaller with time.  In total, after the blue loop the
star has $\sim 5$ times less angular momentum than before, and thus
red supergiants which underwent a blue loop rotate significantly
slower than those which did not.

\section{Discussion}
\lSect{dis} 

The mechanism to spin-up stellar envelopes presented above enhances
the specific angular momentum in the surface layers of the star by
some factor (cf. \Sect{res}); a question to be considered separately
is the origin of the angular momentum, and also in which cases
critical rotation is reached.  For single stars, the available angular
momentum
seems to be
limited to that of the ZAMS star, reduced by the angular momentum lost
during its evolution.  A way to supply a considerable amount of
angular momentum to the star would be the capture of a companion star
or planet (Soker 1996; Podsiadlowski 1998).  This might occur when the
star first becomes a red supergiant. The resulting configuration would
have a much higher angular momentum than a normal single star, but
still, due to the high moment of inertia of the supergiant, it could
initially be far away from critical rotation.  Due to the spin-up
mechanism described here, such a star might then approach critical
rotation even before the transition into a blue supergiant.

The mechanism to spin-up stellar envelopes (\Sect{mech}), which was
applied to a post-red supergiant $12\,\Msun$ star in \Sect{num}, may
operate in all evolutionary phases during which stars evolve from the
Hayashi-line towards the hotter part of the HR diagram. These comprise
the transition of pre-main sequence stars form their fully convective
stage to the main sequence, the transition of low and intermediate
mass stars from the Asymptotic Giant Branch (AGB) to central stars of
planetary nebulae, the blue loops of stars in the initial mass range
$\sim 3\,\mso ... \sim 25\,\mso$, and the transition of massive red
supergiants into Wolf-Rayet (WR) stars. For all four phases,
observational evidence for axisymmetric circumstellar matter exists.

A high value of $\Omega$ during the evolution off the Hayashi line can
be expected to have notable influence not only on the mass and angular
momentum loss rate as in the case studied in \Sect{num}, but also on
the geometry of the wind flow.  At present, theoretical predictions of
the latitudinal dependence of stellar wind properties for rapidly
rotating stars are ambiguous.  Bjorkman \& Cassinelli (1993) have
proposed that angular momentum conservation of particles in a pressure
free wind which is driven by purely radial forces leads, for
sufficiently large values of $\Omega$, to high ratios of equatorial to
polar wind density, i.e. to disks or disk-like configurations.  Owocki
et al. (1996) found that this might no longer be the case when the
non-radial forces occurring in hot star winds are accounted for.  For
cooler stars, the prospects of disk formation may thus be better
(cf. also Ignace et al., 1996). To discus the geometry of the winds of
our models is beyond the scope of this paper; however, we want to
point out that the maximum values of $\Omega$ and $\Jdot/\Mdot$ occur
at about $\Teff\simeq 6\,000\,\K$ (cf. \Sect{res} and \Fig{t-xx}), so
that an equator-to-pole wind density ratio larger than one may still
be justified.

\subsection{Blue loops during or after core helium burning}
\lSect{blue-loop}

An interesting example where the spin-up mechanism described in this
paper should almost certainly have played a r\^{o}le is the progenitor
of SN~1987A. In fact, the SN~1987A progenitor is the only star of
which we are reasonably sure that it performed a blue loop, in this
case after core helium exhaustion (cf. Arnett et al. 1989)

The structures observed around SN~1987A seem to be rotationally
symmetric with a common symmetry axis, which may suggest that rotation
has played a major r\^{o}le in their formation.  The inner of the
three rings is currently explained by the interaction of the blue
supergiant wind with the wind of the red supergiant precursor (cf.,
Chevalier 1996). However, this interaction would result in a spherical
shell in case of spherically symmetric winds. To understand the ring
structure of the inner interaction region, and maybe also the two
outer rings, significant rotation appears therefore to be required
(cf.  Martin \& Arnett 1995), which may be provided by our spin-up
mechanism.  We may note that for this mechanism to work it is
insignificant what actually triggered the blue loop; i.e. it would
work as well for single star scenarios with a final blue loop
(cf. Langer 1991b; Meyer 1997; Woosley et al. 1998) or for binary
merger scenarios which predict a final red $\to$ blue transition of
the merger star (cf. Podsiadlowski 1998).

We do not expect to find ring nebulae frequently around blue
supergiants (cf. Brandner et al. 1997), since the time scale on which
our model (cf. \Sect{num}) shows very high surface rotation rates is
rather small compared to the typical life time of a blue supergiant
(cf. Langer 1991a), and because they are quickly dissolved by the blue
supergiant wind.  SN~1987A is an exception, since here the transition
to the blue happened only short time ago (i.e. the supernova exploded
only shortly after the transition). However, a certain type of
B~supergiants, the {\Be}~stars, show emission line features which
might be due to a circumstellar disk (cf. Zickgraf et al. 1996). The
location of the less luminous subgroup of {\Be}~stars in the HR
diagram (Gummersbach et al. 1995) is in fact consistent with a blue
loop scenario for their evolution.  I.e., their disks might be
produced by the spin-up mechanism described here (cf. also Langer \&
Heger 1998).

\subsection{Red supergiant $\to$ Wolf-Rayet star transition}
\lSect{RSG-WR}

The transition of massive mass losing red supergiants into Wolf-Rayet
stars is the massive star analogue of the AGB $\to$ post-AGB star
transition. It occurs when the mass of the hydrogen-rich envelope is
reduced below a critical value (cf., e.g., Schaller et al. 1992) and
should not be confused with the blue loops discussed in
\Sect{blue-loop}. As in the post-AGB case (\Sect{post-AGB}), the
spin-up mechanism can be expected to work.  Unlike in the case
discussed in \Sect{res}, the major part of the envelope is lost due to
stellar winds before the red supergiant $\to$ Wolf-Rayet star
transition, with the consequence of considerable angular momentum
loss.  Therefore, it may be more difficult for the star to reach
critical rotation during the contraction.

However, there are signs of asphericity in the ring nebula NGC~6888
around the Galactic Wolf-Rayet star WR~136 which has been interpreted
as swept-up red supergiant wind shell by Garc\'{\i}a-Segura \& Mac~Low
(1995) and Garc\'{\i}a-Segura et al. (1996).  Also, Oudmaijer et
al. (1994) report on bi-polar outflows from IRC+10420, a massive star
just undergoing the red-supergiant $\to$ Wolf-Rayet star transition.
IRC+10420 is currently an F~type star, i.e. it has an effective
temperature at which we expect the maximum effect of the spin-up
mechanism discussed here (cf. \Sect{res}). Therefore, in the absence
of a binary companion, the spin-up mechanism may yield the most
promising explanation for the bipolar flows.

\subsection{Pre-main sequence evolution}
\lSect{pre-MS}

As mentioned in \Sect{res}, we started our $12\,\Msun$ sequence from a
fully convective pre-main sequence configuration.  We expected the
spin-up mechanism found for the post-red supergiant stage to be also
present in the contraction phase towards the main sequence. An
analysis of this evolutionary phase showed in fact its presence,
although the efficiency of the spin-up was found to be somewhat
smaller.  During that part of the pre-main sequence contraction phase
where the convective region retreats from the center of the star to
its surface, the star would have increased its equatorial rotational
velocity by a factor of $\sim 3$ if angular momentum were conserved
locally, but it was spun-up by a factor of $\sim 10$. I.e., the
spin-up mechanism described in \Sect{mech} resulted in an additional
increase of the rotational velocity of a factor of $\sim 3$. This was
not enough to bring the star to critical rotation in this phase.
However, if the initial rotation rate would have been larger, a phase
of critical rotation during the pre-main sequence stage would well be
possible.

While a fully convective pre-main sequence stage may not be realistic
for massive stars and is just assumed in our case for mathematical
convenience, pre-main sequence stars of low and intermediate mass are
supposed to evolve through a fully convective stage (Palla \& Stahler
1991).  During the transition from this stage to the main sequence,
the spin-up mechanism described in this paper might operate.  Pre-main
sequence stars in the corresponding phase, i.e. past the fully
convective stage but prior to core hydrogen ignition, are often found
to have disks. They correspond to the T~Tauri stars at low mass (e.g.,
Koerner \& Sargent 1995) and to central stars of Herbig Haro objects
at intermediate mass (e.g., Marti et al. 1993). However, the disks are
usually interpreted as remnants of the accretion process which built
up the star.  Since pre-main sequence stars are often found to be
rapid rotators (cf. Walker 1990), we may speculate here about a
possible contribution to the disk due to decretion from the star
reaching critical rotation due to spin-up (cf. Krishnamurtihi et
al. 1997).

\subsection{Post-AGB evolution}
\lSect{post-AGB}

Low and intermediate mass stars leave the AGB when the mass of their
hydrogen-rich envelope decreases below a certain value.  When this
happens, the envelope deflates, the stellar radius decreases, and the
energy transport in the envelope changes from convective to
radiative. The spin-up mechanism described in \Sect{mech} can be
expected to operate in this situation, as in the red supergiant $\to$
Wolf-Rayet star transition (\Sect{RSG-WR}).

Whether or not post-AGB stars can reach critical rotation due to this
spin-up is not clear.  Certainly, the heavy mass loss during the
evolution on the AGB spins the envelope down according to the
mechanism sketched in \Fig{panels}.  The ratio of the rotation rate to
the critical rotation rate $\Omega$ may be further affected by random
kicks due to asymmetric mass loss, which may keep the envelope at some
level of rotation (cf. Spruit 1998), or by transport of angular
momentum from the core to the envelope, which may be efficient during
the thermal pulses, and by the evolution of the critical rotation
rate, which depends on the Eddington factor $\Gamma$ and thereby on
the opacity coefficient (cf. Garc\'{\i}a-Segura et al. 1998).

Clearly, the spin-up mechanism described in this paper may help to
bring post-AGB stars --- or more specific: stars which just left the
AGB, i.e. central stars of proto-planetary nebulae --- closer to
critical rotation.  It may play a r\^{o}le in explaining axisymmetric
flows which are often observed in central stars of proto-planetary
nebulae (cf. Kwok 1993) and the shapes of bi-polar planetary nebulae
(cf. Schwarz et al. 1993; Stanghellini et al. 1993; Garc\'{\i}a-Segura
et al. 1998).

\section{Summary and conclusions}
\lSect{con}

In this paper, we discussed the effect of mass outflow through the
inner or outer boundary of a rigidly rotating envelope on its rotation
frequency.  It causes a change of the specific angular momentum in the
envelope and alters its rotation rate besides what results from
contraction or expansion (cf. \Fig{t-v}). For constant upper and lower
boundaries of the envelope, which we found a good approximation for
convective envelopes (cf. \Fig{t-r-m}), a spin-down occurs for mass
outflow through the upper boundary --- which corresponds, e.g., to the
case of stellar wind mass loss from a convective envelope (cf. also
Langer \mbox{1998) ---,} while a spin-up results from mass outflow
through the lower boundary (cf. \Fig{panels}).

The latter situation is found in evolutionary models of a rotating
$12\,\Msun$ star at the transition from the Hayashi-line to the blue
supergiant stage.  The star increased its rotational velocity one
order of magnitude above the velocity which would result in the case
of local angular momentum conservation.  It would have increased its
rotational velocity even further if it would not have arrived at
critical rotation (cf. \Sect{res}, \Fig{t-v}), with the consequence of
strong mass and angular momentum loss.  At this point, the specific
angular momentum loss $\Jdot/\Mdot$ reached about $8\,10^{19}\,\junit$
(cf. \Fig{t-xx}).

The geometry of circumstellar matter around stars which undergo a red
$\to$ blue transition may be strongly affected by the spin-up.  We
propose that this was the case for the progenitor of SN~1987A, the
only star of which we know that it performed a red $\to$ blue
transition in the recent past. The blue supergiant in its neighborhood
studied by Brandner et al. (1997), around which they found a ring
nebula as well, is another candidate.  Also, {\Be}~stars may be
related with the post-red supergiant spin-up
(cf. \Sect{blue-loop}). Furthermore, the spin-up mechanism studied in
this paper may be relevant for bipolar outflows from central stars of
proto-planetary nebulae (\Sect{post-AGB}), from stars in the
transition phase from the red supergiant stage to the Wolf-Rayet stage
(\Sect{RSG-WR}), and from pre-main sequence stars (\Sect{pre-MS}).

\begin{acknowledgements}
  We are grateful to S.E. Woosley and S.P. Owocki for helpful
  discussions, to J.~Fliegner providing us with an implementation of
  rotational physics in the stellar evolution code, and to A. Weiss
  for supplying us with his recent opacity tables. This work has been
  supported by the Deutsche Forschungsgemeinschaft through grant
  No. La~587/15-1.
\end{acknowledgements}

\end{document}